\documentclass[aps,prd,superscriptaddress,nofootinbib,preprintnumbers,12pt]{revtex4}

\usepackage{amsmath}
\usepackage{graphicx}
\usepackage{dcolumn}
\usepackage{bm}
\usepackage{amssymb}
\usepackage{latexsym}

\newcommand{\be}{\begin{equation}}
\newcommand{\ee}{\end{equation}}
\newcommand{\bq}{\begin{eqnarray}}
\newcommand{\eq}{\end{eqnarray}}
\begin{document}

\title{Brane inflation revisited after WMAP five-year results}
\author{Yin-Zhe Ma}
\email{yzm20@cam.ac.uk}
\affiliation{Institute of Astronomy,
University of Cambridge, Madingley Road, Cambridge CB3 0HA, United
Kingdom}
\author{Xin Zhang}
\email{zhangxin@mail.neu.edu.cn}
\affiliation{College of Sciences,
Northeastern University, Shenyang 110004, Liaoning, People's
Republic of China}

\begin{abstract}
In this paper, we revisit brane inflation models with the WMAP
five-year results. The WMAP five-year data favor a red-tilted power
spectrum of primordial fluctuations at the level of two standard
deviations, which is the same as the WMAP three-year result
qualitatively, but quantitatively the spectral index is slightly
greater than the three-year value. This result can bring impacts on
brane inflation models. According to the WMAP five-year data, we
find that the KKLMMT model can survive at the level of one standard
deviation, and the fine-tuning of the parameter $\beta$ can be
alleviated to a certain extent at the level of two standard
deviations.
\end{abstract}

\maketitle

\section{Introduction}
The inflation paradigm of the early universe provides a compelling
explanation for many big puzzles in the standard big bang cosmology,
such as the problems of homogeneity, isotropy and flatness of the
universe \cite{Guth81}. This period of accelerated expansion in the
early universe predicts a nearly scale-invariant perturbation power
spectrum, which has already been supported by the measurement of
temperature fluctuations in the cosmic microwave background
radiation \cite{Smoot92,Bennett03}. In spite of many
phenomenological successes of the inflation based on effective field
theory, there exist serious problems, concerning the origin of the
scalar field driving inflation, namely the singularity problem and
the trans-Plankian problem. Therefore, it is expected that inflation
should be realized in a more natural way from some fundamental
theory of microscopic physics. String theory is one of the most
promising candidates for the fundamental microscopic physics, so it
is very important to try to embed the inflation scenarios into
string theory. One possible inflation scenario within the framework
of string theory is the brane inflation, in which the inflation is
driven by the potential between the parallel dynamic brane and
anti-brane \cite{Dvali98,Burgess01,Quevedo02}. The distance between
the branes in the extra compactified dimensions plays the role of
the inflaton field. However, generically, getting a sufficiently
flat inflaton potential in brane inflation models is not an easy
thing \cite{Quevedo02}.

A rather realistic slow roll inflation scenario based on string
theory was first proposed by Kachru et al. \cite{kklt}. In
\cite{kklt}, by introducing some $\overline{\rm D}$3-branes in a
warped geometry in type IIB superstring theory to break
supersymmetry, the authors successfully lift the AdS vacuum to a
metastable de Sitter vacuum with sufficiently long lifetime. This
mechanism is often called KKLT mechanism. Furthermore, by putting an
extra pair of the brane and anti-brane in this scenario, one can
successfully realize a slow roll inflation model, namely the KKLMMT
model \cite{Kachru03}. In this model, the anti-brane is fixed at the
bottom of a warped throat, while the brane is mobile and experiences
a small attractive force towards the anti-brane. The inflation ends
when the brane and the anti-brane collide and annihilate, initiating
the hot big bang epoch. The annihilation of the brane and anti-brane
makes the universe settle down to the string vacuum state that
describes our universe. During the process of the brane collision,
cosmic strings are copiously produced \cite{Jones02,Sarangi02}. For
a well-worked inflationary scenario, the production of topological
defects other than cosmic strings must be suppressed by many orders
of magnitude. For extensive studies on the KKLMMT model and other
types of brane inflation models, see, e.g.,
\cite{Tye05,Baumann06,Huang06,Huang:2006zu,Zhang06,Bean:2007hc,Lorenz:2007ze,Alabidi:2008ej}.

Recently, the WMAP team released the five-year data
\cite{Komastu08}. For the $\Lambda $CDM model, WMAP five-year data
show that the index of power spectrum satisfies\footnote{The pivot
scale is taken to be $k_{\rm pivot}=0.002 {\rm Mpc}^{-1}$.}
\begin{equation}
n_{s}=0.963_{-0.015-0.028}^{+0.014+0.029}~~~(1\sigma, 2\sigma~
\text{CL});  \label{nsnumber1}
\end{equation}
combining WMAP with SDSS and SNIa, the result is
\begin{equation}
n_{s}=0.960_{-0.013-0.027}^{+0.014+0.026}~~~(1\sigma, 2\sigma~
\text{CL}),  \label{nsnumber2}
\end{equation}
which is a little bluer than the WMAP three-year result, though the
red power spectrum is still favored at the level of $3\sigma $ CL.
Meanwhile, the running of the spectral index is not favored anymore.
With WMAP five-year data only, the running of the spectral index
is\footnote{In this case, the result of the spectral index is
$n_s=1.031^{+0.054}_{-0.055}$.}
\begin{equation}
\alpha _{s}=\frac{dn_{s}}{d\ln
k}=-0.037_{-0.028}^{+0.028}~~~(1\sigma~\text{CL}).
\label{runnumber1}
\end{equation}
combining with SDSS and SNIa data, the result is\footnote{In this
case, the result of the spectral index is
$n_s=1.022^{+0.043}_{-0.042}$.}
\begin{equation}
\alpha _{s}=\frac{dn_{s}}{d\ln
k}=-0.032_{-0.020}^{+0.021}~~~(1\sigma~\text{CL}).
\label{runnumber2}
\end{equation}
It is notable that the result of the spectral index in five-year
WMAP is slightly larger than that in three-year WMAP. More
evidently, the result of the running of the spectral index in
five-year WMAP is much smaller than that in three-year WMAP.
Finally, let's see the result of the amplitude of the primordial
gravitational waves. When combining WMAP data with SDSS and SNIa
data, the tensor-to-scalar ratio $r$ is limited under a much tighter
bound:\footnote{In this case, the result of the spectral index is
$n_s=0.968\pm 0.015$.}
\begin{equation}
r<0.20~~~(95\%~\text{CL}).  \label{tensornumber1}
\end{equation}

Based on the WMAP three-year results, Huang et al. \cite{Huang06}
investigated brane inflation models and showed that the KKLMMT
model cannot fit WMAP+SDSS data at the level of one standard
deviation and a fine-tuning (at least eight parts in a thousand)
is needed at the level of two standard deviations. Now, the WMAP
data are updated to the five-year ones, so it is meaningful to see
how the status of brane inflation is affected by the arrival of
the new WMAP data. Therefore, in this paper, we address this
issue, i.e., we revisit brane inflation with the WMAP five-year
results.

This paper is organized as follows: In section \ref{sec:Dp}, we
discuss a simple brane inflation model neglecting the problem of
dynamic stabilization. In section \ref{sec:kklmmt}, we focus on
the KKLMMT model and investigate to what degree the model is
fine-tuned under the WMAP five-year data. The conclusion is given
in the last section.
\section{A general brane Inflation model}\label{sec:Dp}
First, let's consider some general brane inflation models. Consider
a pair of $Dp$ and $\bar{D}p$-branes ($p\geq 3$) filling the four
large dimensions and separated from each other in the extra six
dimensions that are compactified. The inflaton potential is given by
\cite{Quevedo02,Huang06}
\begin{equation}
V=V_{0}\left(1-\frac{\mu^n }{\phi^n }\right),  \label{potential1}
\end{equation}
where $V_0$ is an effective cosmological constant on the brane
provided by brane tension, and the second term in (\ref{potential1})
comes from the attractive force between the branes. The parameter
$n$ satisfies the relation $n=d_{\perp}-2$, where $d_{\perp}=9-p$ is
the number of transverse dimensions. Obviously, we have $n\leq 4$
due to $d_{\perp}\leq 6$. Corresponding to the $e$-folding number
$N$ before the end of inflation, the inflaton field $\phi$ takes the
value
\begin{equation}
\phi _{N}=[N M_{pl}^{2}\mu ^{n}n(n+2)]^{1/(n+2)},  \label{begin1}
\end{equation}
where $M_{pl}$ is the reduced Planck mass of four dimensional world.
The slow-roll parameters can be given as
\begin{equation}
\epsilon
_{v}=\frac{M_{pl}^{2}}{2}\left(\frac{V'}{V}\right)^{2}=\frac{n^{2}
}{2(n(n+2))^{\frac{2(n+1)}{n+2}}}\left(\frac{\mu
}{M_{pl}}\right)^{\frac{2n}{n+2}}N^{- \frac{2(n+1)}{n+2}},
\label{epsilon1}
\end{equation}%
\begin{equation}
\eta _{v}=M_{pl}^{2}\frac{V''}{V}=-\frac{n+1}{n+2}\frac{1}{N},
\label{eta1}
\end{equation}%
\begin{equation}
\xi
_{v}=M_{pl}^{4}\frac{V'V'''}{V^{2}}=\frac{n+1}{n+2}\frac{1}{N^{2}}.
\label{xi1}
\end{equation}%
Due to the fact that $\mu$ is much less than $M_{pl}$, $\xi_v$ is
negligible. Hence, the amplitude of the tensor perturbations is
negligible \cite{Huang06}. One then obtains the spectral index and
its running:
\begin{equation}
n_{s}=1-\frac{n+1}{n+2}\frac{2}{N},  \label{power1}
\end{equation}%
\begin{equation}
\alpha_{s}=-\frac{n+1}{n+2}\frac{2}{N^{2}}.  \label{running1}
\end{equation}%
It is easy to see that the running of the spectral index $\alpha_s$
is also negligible.
\begin{figure}[tbh]
\centerline{\includegraphics[bb=0 0 434 291,
width=3.3in]{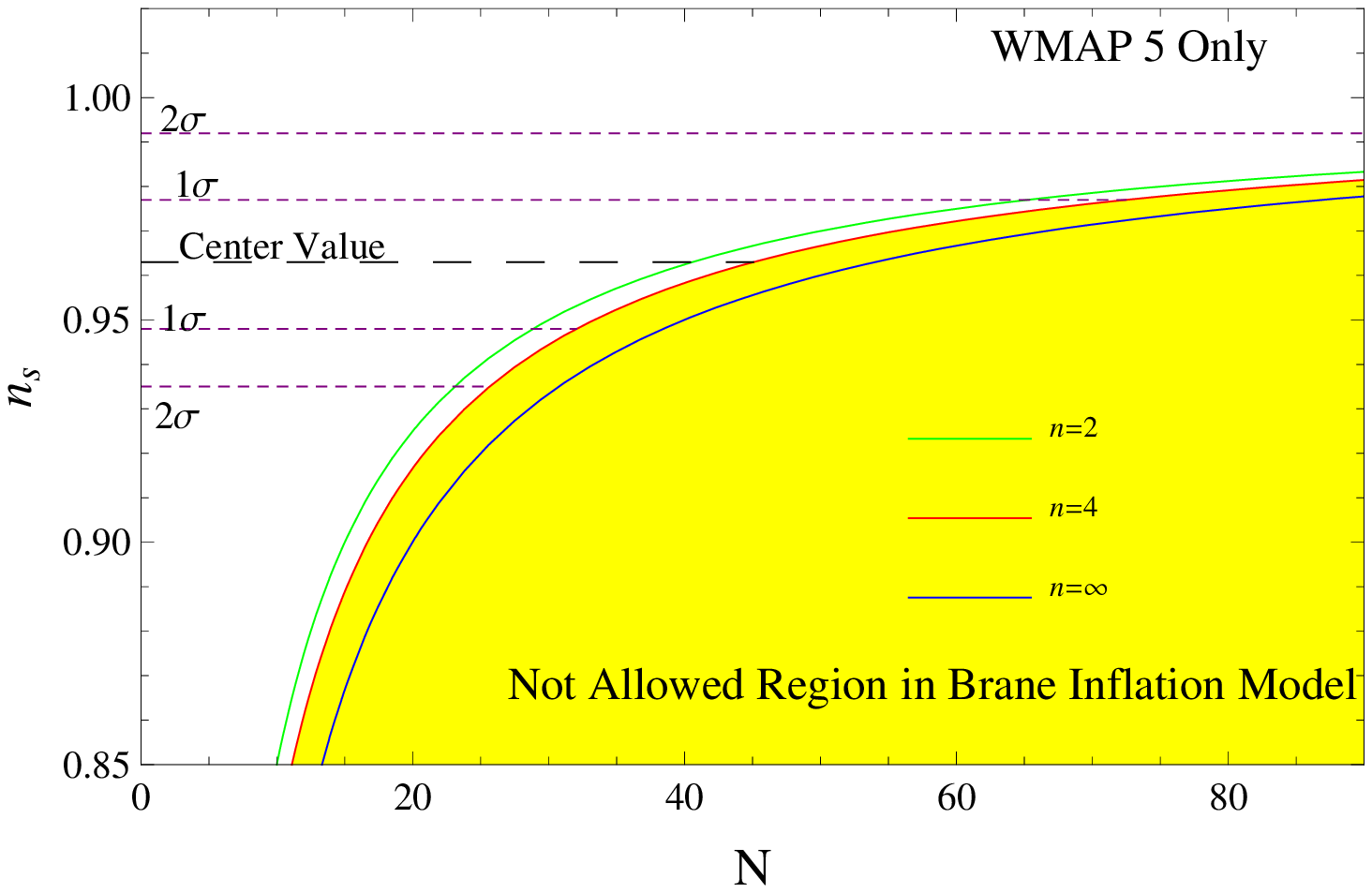}
\includegraphics[bb=0 0 440 287,width=3.3in]{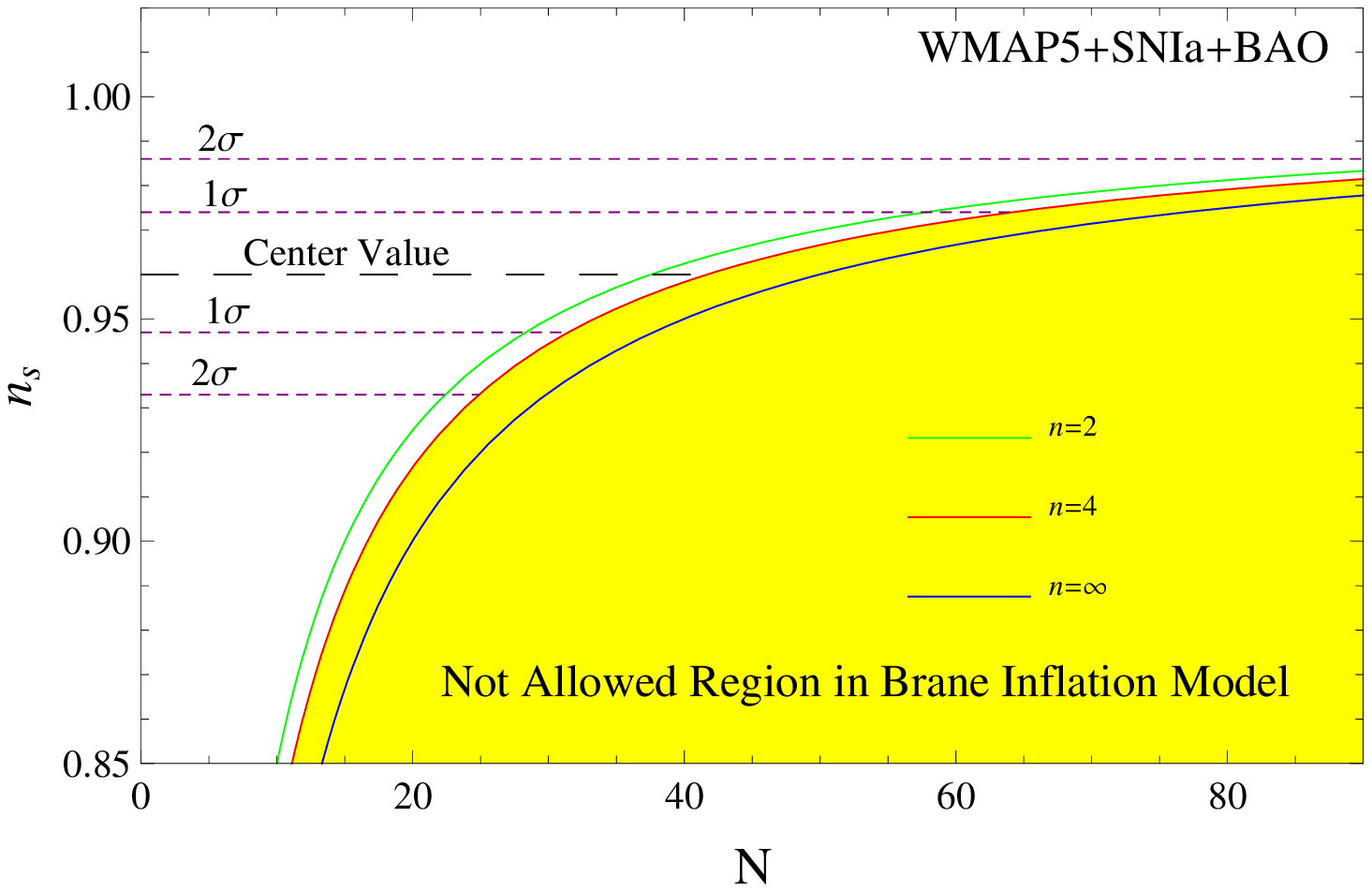}}
\caption{\small Comparison of the WMAP five-year results of the
special index with the brane inflation model. Note that since the
number of the brane transverse dimensions is not greater than $6$,
the region with $n>4$ is not allowed (yellow region). Left: WMAP
five-year results only; right: WMAP five-year data combined with
SNIa and BAO data. } \label{ns1}
\end{figure}

Figure \ref{ns1} shows the spectral index for different parameter
$n$ and $e$-folding number $N$. The region of $n>4$ is not allowed
by theory, since the number of transverse dimensions is less than
six. We focus on the two cases of $n=2$ and 4 corresponding
respectively to the $D5$ and $D3$-brane cases. From Fig. \ref{ns1},
we see that the brane inflation model is consistent with the WMAP
data within some range of $N$. On the other hand, it has been
proposed in \cite{Alabidi06} that the $e$-folding number relevant to
observations should be
\begin{equation}
N=54\pm 7.  \label{e-folds}
\end{equation}
So, we shall check whether the fitting of WMAP five-year data can
give a reasonable $e$-folding number consistent with the result of
(\ref{e-folds}), namely, $47<N<61$.
\begin{table}[tbp]
\begin{centering}
\caption{Fitting results for the number of $e$-folds $N$}
\label{tab1}
\begin{tabular}{|c|c|c|c|c|}\hline
{Model} & WMAP5 Only & WMAP5+SNIa+BAO &WMAP5+ACBAR
&WMAP5+CMB\footnote{This includes the small-scale CMB measurements
from CBI, VSA, ACBAR and BOOMERanG, see
\cite{Komastu08,Lambda,CMBdata}.}
\\\hline $D5$-brane ($n=2$) & $N=40.5^{+24.7+147.0}_{-29.4-23.1}$ &
$N=37.5^{+20.2+69.6}_{-9.2-15.1}$&$N=41.7^{+26.5+145.8}_{-11.7-17.9}$&$N=37.5^{+20.2+77.9}_{-9.7-15.1}
$
 \\ \hline $D3$-brane ($n=4$) & $N=45.0^{+27.5+163.3}_{-12.9-19.4}$ &
$N=41.7^{+22.4+77.3}_{-10.3-16.8}$&$N=46.3^{+29.5+162.0}_{-13.0-19.8}$&$N=41.7^{+22.4+86.5}_{-10.8-16.8}
$
\\\hline
\end{tabular}
\end{centering}
\end{table}

Table \ref{tab1} is the result of the fitting to the WMAP data
(combined with other observational data) for the number of $e$-folds
$N$. It is clear to see that the fitting values of the $e$-folding
number are generally consistent with the result of \cite{Alabidi06},
$N=47\sim 61$, within $1\sigma $ range. Therefore, testing with the
WMAP five-year data, one finds that the model survives at the level
of 1$\sigma$.

\begin{figure}[tbh]
\centerline{\includegraphics[bb=0 0 441 295,
width=3.3in]{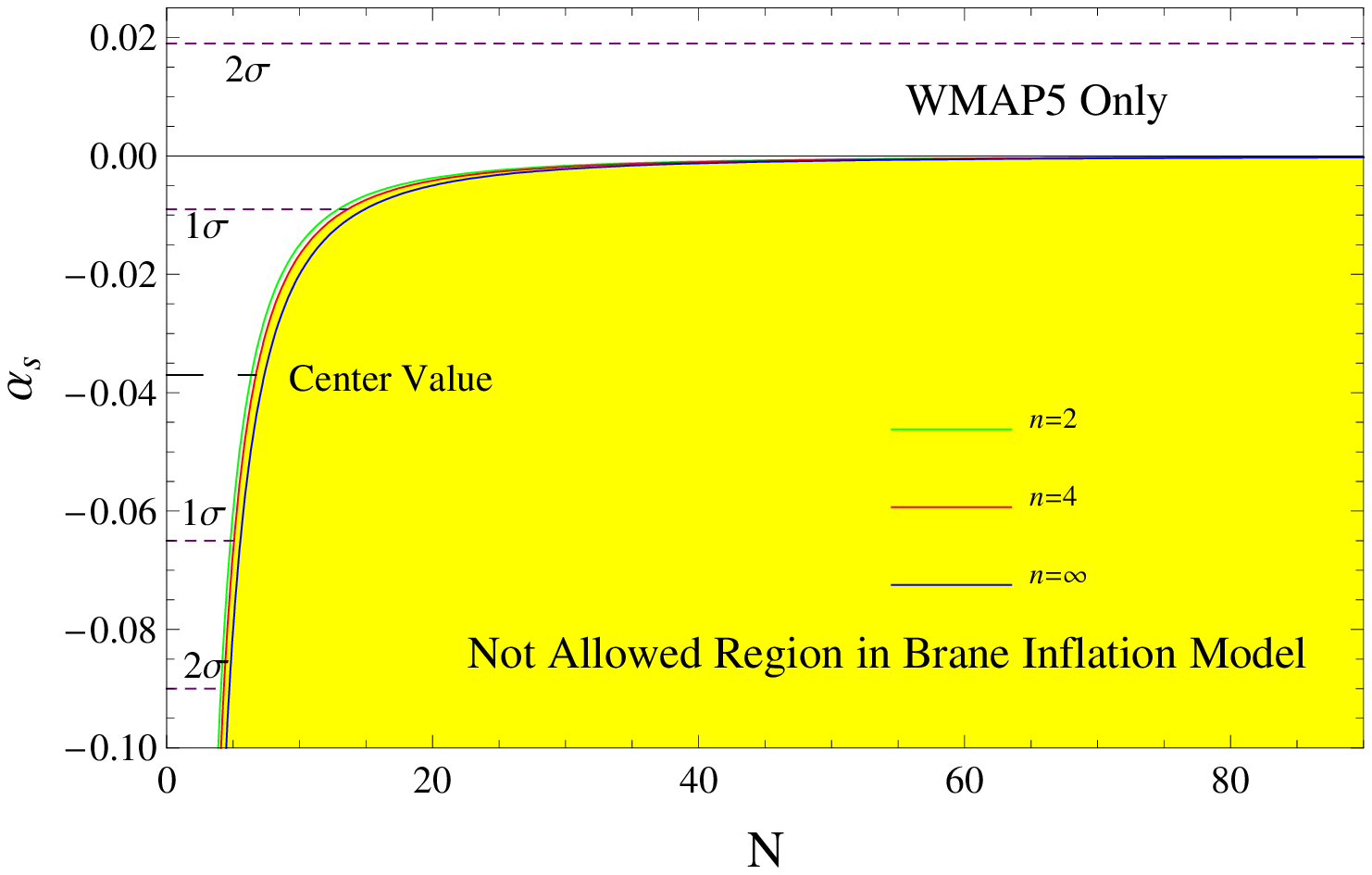}
\includegraphics[bb=0 0 447 297,width=3.3in]{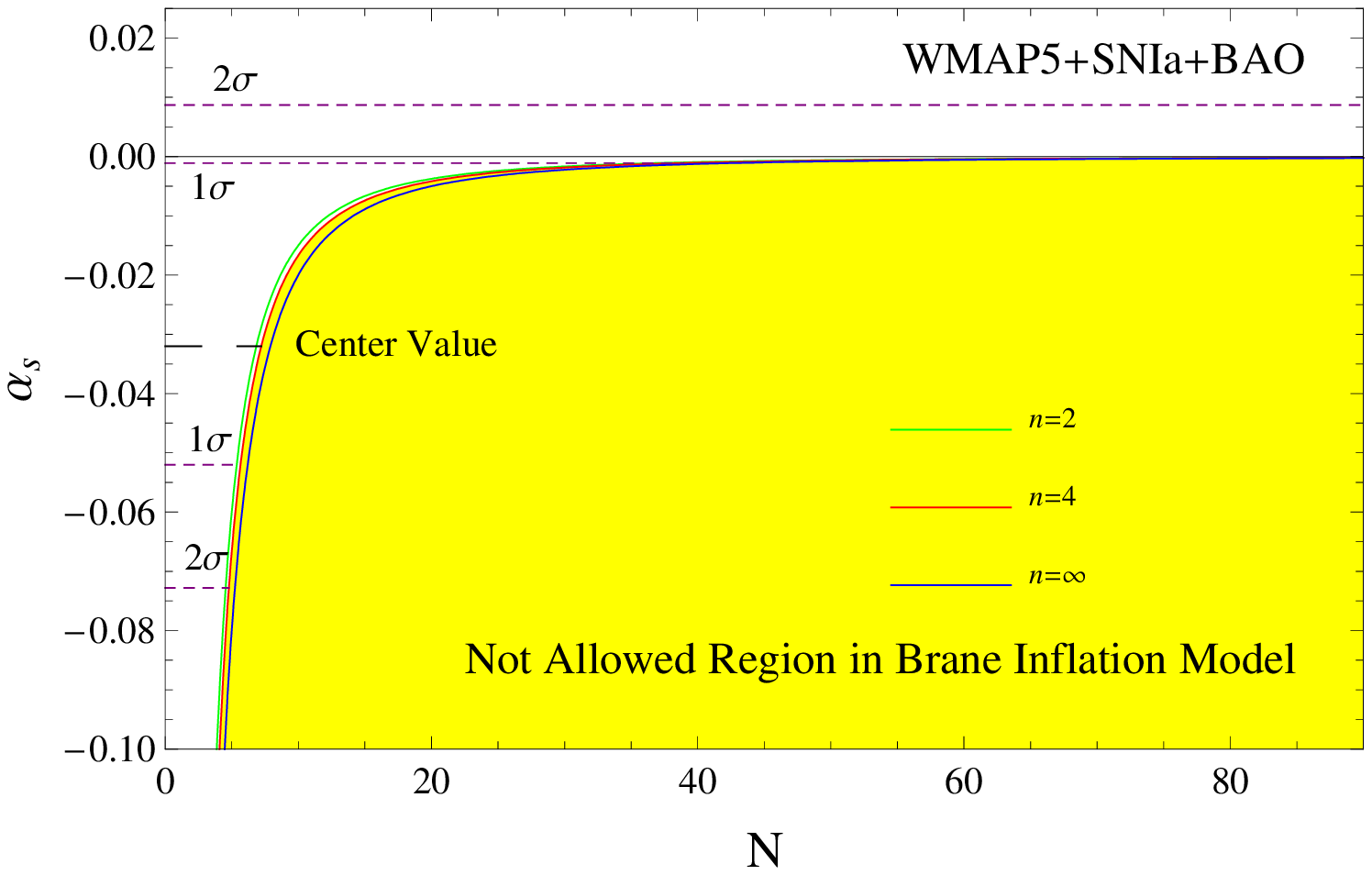}}
\caption{Comparison of the WMAP five-year results of the running of
special index with the brane inflation model. The region with $n>4$
is not allowed (yellow region). Left: WMAP five-year results only;
right: WMAP five-year data combined with SNIa and BAO data. }
\label{as1}
\end{figure}

\begin{figure}[tbh]
\centerline{\includegraphics[bb=-100 194 386
608,scale=0.5]{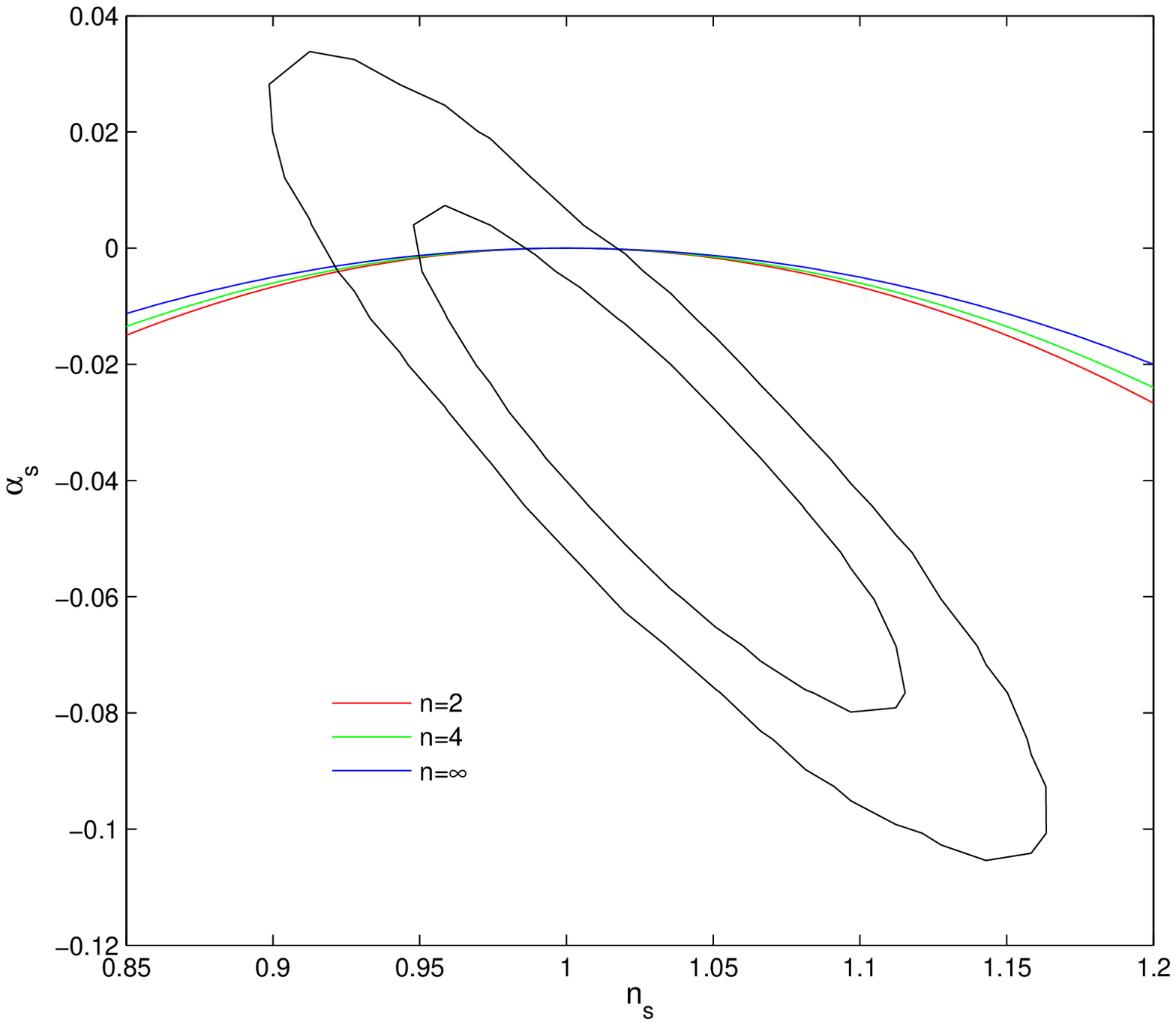}} \caption{Comparing the brane inflation
model to the WMAP five-year data (WMAP5 only): the running of the
spectral index.} \label{model1}
\end{figure}

Furthermore, let us consider the running of the spectral index. We
plot $\alpha_s$ vs. $N$ in Fig. \ref{as1}. From the left panel of
this figure, we find that though the brane inflation model cannot
yield the right results of the running of the spectral index given
by WMAP five-year data in 1$\sigma$ range with a reasonable
$e$-folding number, but it can be consistent with the
observational results in 2$\sigma$ range. From the right panel of
Fig. \ref{as1}, one finds that, for WMAP5+SN+BAO, the model is
even consistent with the data in 1$\sigma$ range. Fig.
\ref{model1} shows the joint two-dimensional marginalized
distribution of the spectral index $n_s$ and its running
$\alpha_s$, for WMAP five-year data. From this figure, we see that
the trajectories of the brane inflation model (D5 and D3-brane
cases) are within the region of 1$\sigma$ CL.


\begin{figure}[tbh]
\centerline{\includegraphics[bb=0 0 515 333, width=3.3in]{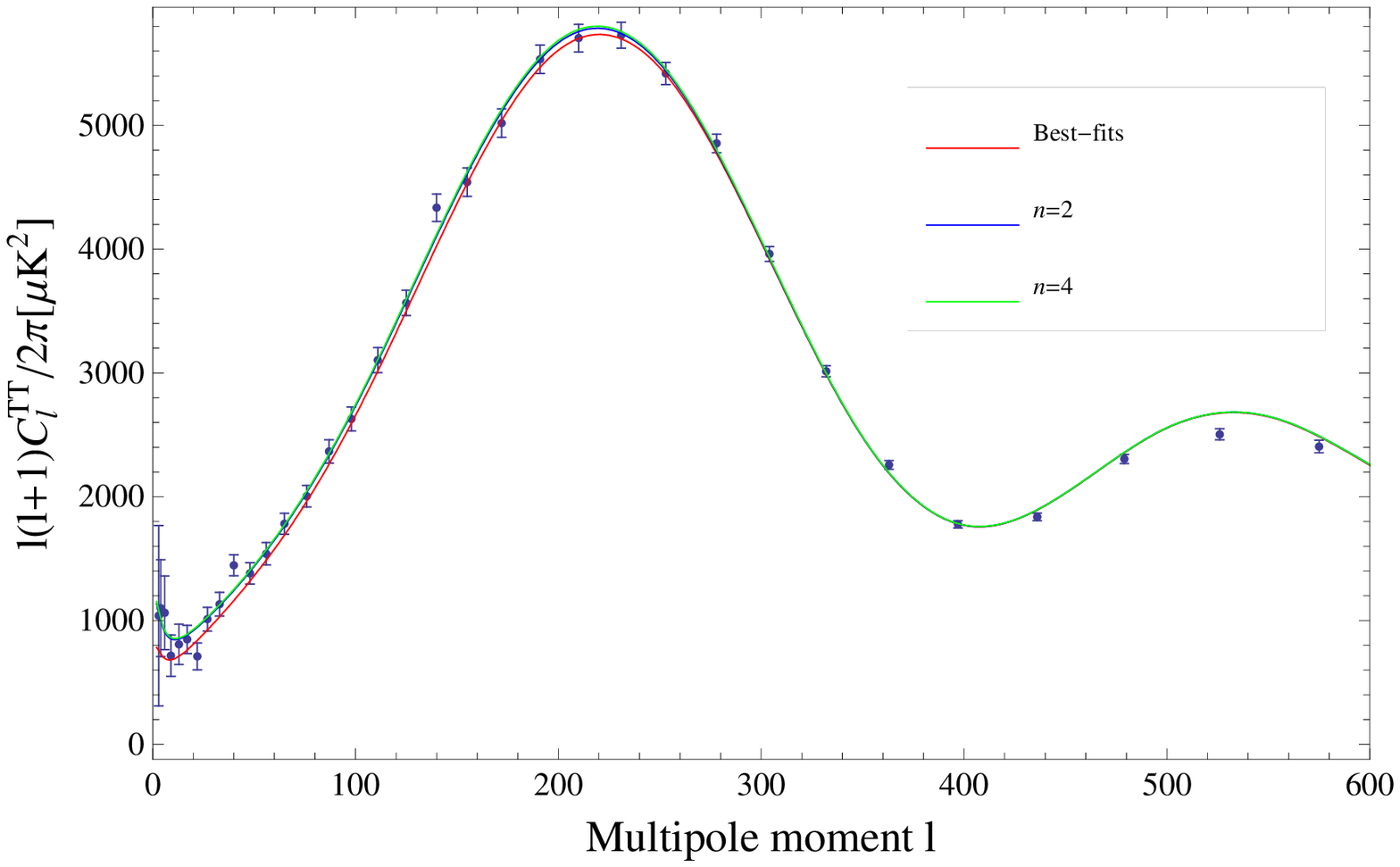}
\includegraphics[bb=0 0 520 332,width=3.3in]{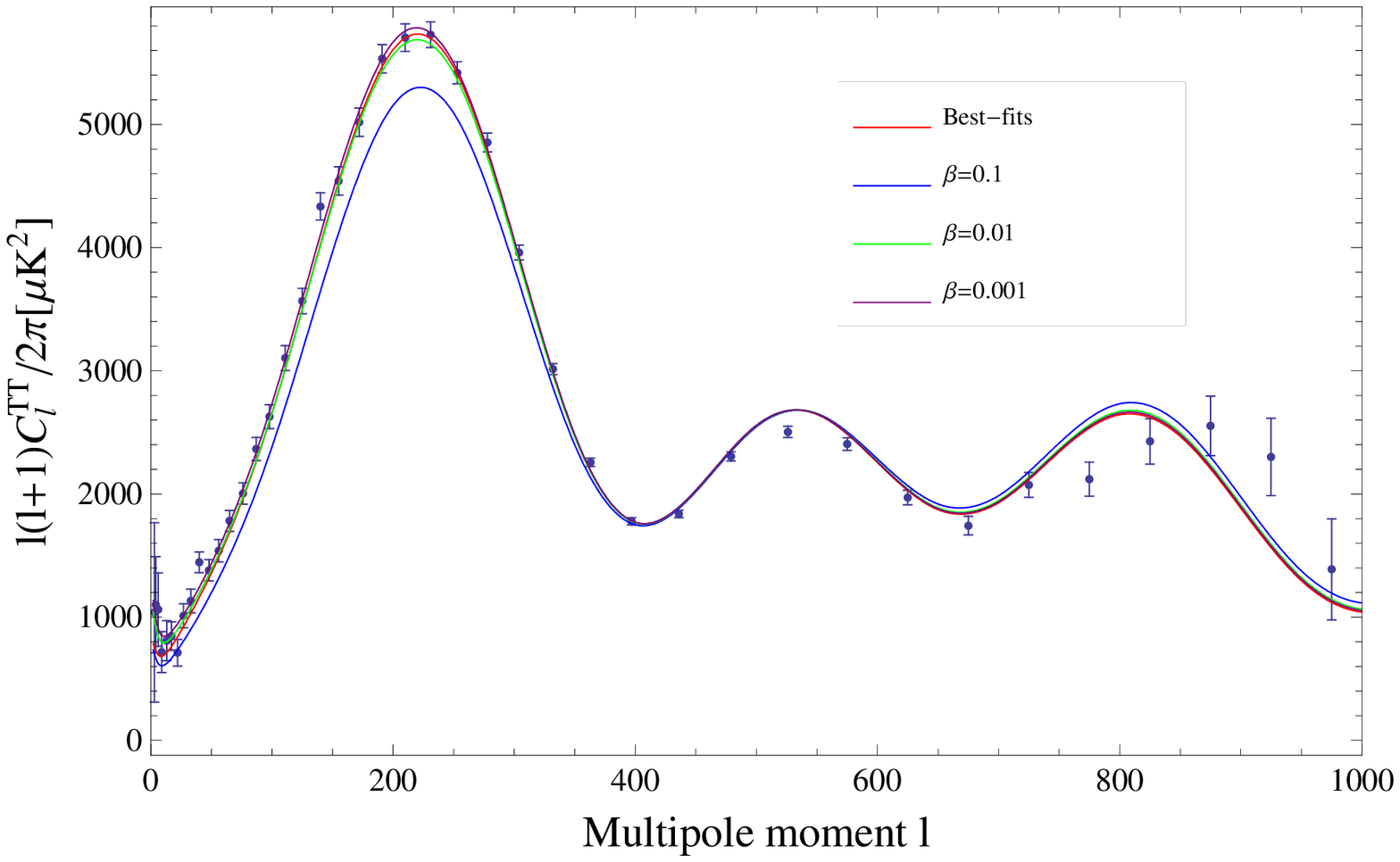}}
\caption{Comparison of angular power spectrum of the brane
inflation models with the WMAP five-year data. Left: the general
brane inflation model. Right: the KKLMMT model.} \label{angular}
\end{figure}

By modifying the CAMB code \cite{CAMB}, we also compare the angular
power spectrum of the model with the WMAP five-year data, see Fig.
\ref{angular}. In the left panel, we see that the angular power
spectrum produced by the model is generally consistent with the WMAP
five-year data, although the cases with $n=2$ and $n=4$ have little
greater amplitudes of the low multipole moments (basically
quadrupole and octopole) compared with the best fit of the WMAP
five-year data. However, due to the cosmic variance and statistical
errors, the low multipole moments have large error bars, the model
prediction is consistent with the observational data.

The above analyses show that the general brane inflation model can
survive confronting the WMAP five-year data. When considering the
WMAP five-year result of the spectral index, the model survives at
the level of 1$\sigma$. This conclusion is different from that of
the analysis of WMAP three year data. It was indicated in
\cite{Huang06} that, with the WMAP three year data, the brane
inflation model for $n=4$ is near the boundary of the WMAP3 only
and cannot fit the WMAP3+SDSS data at the level of 1$\sigma$; the
case with $n=2$ is outside the range allowed by WMAP3 only or
WMAP3+SDSS at the level of 1$\sigma$. So, we find that the WMAP
five-year data save the brane inflation model to the level of
1$\sigma$.

However, the brane inflation model discussed above is not a
realistic model. An important problem is neglected in the above
discussions, which is that the distance between the brane and the
anti-brane must be larger than the size of the extra dimensional
space if the inflation is slow rolling (or $\eta$ is sufficiently
small) in this scenario. Of course, there may exist such a
possibility that the potential in (\ref{potential1}) can emerge in
other theories that we do not know currently, as pointed out in
\cite{Huang06}. The first realistic brane inflation model is the
so-called KKLMMT model \cite{Kachru03} that will be discussed in the
subsequent section.
\section{The KKLMMT model}\label{sec:kklmmt}
The KKLMMT model is set up in the type IIB string theory. In this
model, the highly warped compactifications are considered, with
all moduli stabilized by the combination of fluxes and
non-perturbative effects \cite{Kachru03}. Once a small number of
$\overline{\rm D}$3-branes are added, the vacuum can be
successfully lifted to dS. Furthermore, one can add an extra pair
of D3-bane and $\overline{\rm D}$3-brane in a warped throat, with
the D3-brane moving towards the $\overline{\rm D}$3-brane that is
located at the bottom of the throat. The scenario of brane
inflation can thus be realized. The warped geometry successfully
solves the $\eta$ problem, i.e., it can provide a natural
mechanism for achieving a sufficiently flat potential. In this
section, we shall discuss the KKLMMT model with the WMAP five-year
results.
\subsection{The setup and theoretical results}
Consider a type IIB orientifold compactified on a Calabi-Yau 3-fold
with fluxes, where all moduli are stabilized \cite{Tye05,Baumann06}.
Inside the bulk of the Calabi-Yau manifold, there are throats with
warped geometry, where the metric has the approximate $AdS_{5}\times
S_{5}$ form, where $X_{5}$ is some orbifold of $S_{5}$ and the
$AdS_{5}$ metric in Poincare coordinates has the form
\cite{Tye05,Baumann06}
\begin{equation}
ds^{2}=h^{-\frac{1}{2}}(r)(-dt^2+a(t)^2d\vec{x}^2)+h^{\frac{1}{2}
}(r)ds_6^2,  \label{warp metric}
\end{equation}
where the $h(r)$ is the warp factor
\begin{equation}
h(r)=\frac{R^4}{r^4},  \label{warped factor}
\end{equation}%
in which $R$ is the radius of curvature of the $AdS_{5}$ throat.
Following \cite{Tye05,Huang06}, we consider the potential for the
KKLMMT model given by
\begin{equation}
V(\phi )=\frac{1}{2}\beta H^{2}\phi ^{2}+2T_{3}h^{4}(1-\frac{\mu ^{4}}{\phi
^{4}}),  \label{V}
\end{equation}%
where $T_{3}$ is the D3-brane tension and we have
the relation $\mu ^{4}=\frac{27}{32\pi ^{2}}T_{3}h^{4}$. Then we have%
\begin{equation}
V(\phi )=\frac{1}{2}\beta H^{2}\phi ^{2}+\frac{64\pi ^{2}\mu ^{4}}{27}(1-%
\frac{\mu ^{4}}{\phi ^{4}}).  \label{V1}
\end{equation}%
So the derivative of this potential is
\begin{equation}
V'(\phi )=\beta H^{2}\phi +\frac{256\pi ^{2}\mu ^{8}}{27}\frac{1%
}{\phi ^{5}}.  \label{V derivative}
\end{equation}%
Under the slow roll approximation, we have the Friedmann equation
\begin{equation}
3M_{pl}^{2}H^{2}\simeq V(\phi )\simeq V_{0}=\frac{64\pi ^{2}\mu
^{4}}{27}, \label{Friedman}
\end{equation}%
which leads to the approximation relation $H^{2}=\frac{64\pi ^{2}\mu
^{4}}{27\times 3M_{pl}^{2}}$ we will use in the following
calculation.

The procedure to obtain the inflaton field value at the moment of
the $e$-folding number $N$ before the end of inflation is as
follows: First, we get the inflaton field value at the end of
inflation by using the condition $\eta =-1$; then, we use the
definition of $e$-folding number $N=\int Hdt$ to obtain the field
value at the moment corresponding to the e-folding number $N$.
Therefore, we first calculate the slow-roll parameter $\eta$:
\begin{equation}
\eta =M_{pl}^{2}(\frac{V''}{V})\simeq M_{pl}^{2}(\frac{V''}{V_{0}}),
\end{equation}
where
\begin{equation}
V''=\beta H^{2}-\frac{1280\pi ^{2}\mu ^{8}}{27}\phi ^{-6}.
\end{equation}
Substituting $H^{2}=\frac{64\pi ^{2}\mu ^{4}}{27\times 3M_{pl}^{2}}$
into the above equation, we obtain
\begin{equation}
\eta =\frac{\beta }{3}-20\mu ^{4}M_{pl}^{2}\frac{1}{\phi ^{6}}.
\label{yita2}
\end{equation}
Thus, the field value at the end of inflation $\phi _{f}$ satisfies
\begin{equation}
-1=\frac{\beta }{3}-20\mu ^{4}M_{pl}^{2}\frac{1}{\phi _{f}^{6}},
\end{equation}%
and we obtain
\begin{equation}
\phi _{f}^{6}=\frac{20\mu ^{4}M_{pl}^{2}}{1+\frac{\beta }{3}}.  \label{phif}
\end{equation}
Under the slow roll approximation, we have
\begin{eqnarray*}
N &=&\int_{\phi _{f}}^{\phi _{N}}dN=(3H^{2})\int_{\phi _{f}}^{\phi _{N}}%
\frac{d\phi }{V'(\phi )} \\
&=&(3H^{2})\int_{\phi _{f}}^{\phi _{N}}\frac{d\phi }{\beta H^{2}\phi +\frac{%
256\pi ^{2}\mu ^{8}}{27}\frac{1}{\phi ^{5}}} \\
&=&\frac{1}{2\beta }\ln \left[256\pi ^{2}\mu ^{8}+27H^{2}\beta \phi
^{6}\right]_{\phi _{f}}^{\phi _{N}}.
\end{eqnarray*}%
Substituting $H^{2}=\frac{64\pi ^{2}\mu ^{4}}{27\times 3M_{pl}^{2}}$
and Eq. (\ref{phif}) into the above equation, we obtain
\begin{eqnarray}
\phi _{N}^{6} &=&e^{2\beta N}\phi _{f}^{6}+\frac{12M_{pl}^{2}}{\beta
}\mu
^{4}(e^{2\beta N}-1)  \notag \\
&=&24M_{pl}^{2}\mu ^{4}[\frac{e^{2\beta N}(1+2\beta )-(1+\frac{1}{3}\beta )}{%
2\beta (1+\frac{1}{3}\beta )}],
\end{eqnarray} or
\begin{equation}
\phi _{N}^{6}=24M_{pl}^{2}\mu ^{4}m(\beta ),  \label{ending2}
\end{equation}
where
\begin{equation}
m(\beta )=\frac{e^{2\beta N}(1+2\beta )-(1+\frac{1}{3}\beta )}{2\beta (1+%
\frac{1}{3}\beta )}.  \label{m(beta)}
\end{equation}

With the value of $\phi_N$, we can write
\begin{equation}
\epsilon _{v}=\frac{1}{18}\left(\frac{\phi
_{N}}{M_{pl}}\right)^{2}\left[\beta +\frac{1}{2m(\beta
)}\right]^{2}, \label{epsilon2}
\end{equation}
\begin{equation}
\eta_v =\frac{\beta }{3}-\frac{5}{6}\frac{1}{m(\beta )},
\label{yita3}
\end{equation}
\begin{equation}
\xi _{v}=\frac{5}{3}\frac{1}{m(\beta )}\left[\beta
+\frac{1}{2m(\beta )}\right]. \label{xi2}
\end{equation}
The amplitude of the scalar comoving curvature fluctuations has been
given in \cite{Tye05,Huang06}:
\begin{equation}
\Delta _{\cal R}^{2}\simeq \frac{V}{M_{pl}^{4}}\frac{1}{24\pi
^{2}\epsilon _{v}}=\frac{2}{27m(\beta )}\left(\beta
+\frac{1}{2m(\beta )}\right)^{-2}\left(\frac{\phi
_{N}}{M_{pl}}\right)^{4}, \label{amplitude}
\end{equation}
thus we have
\begin{equation}
\frac{\phi _{N}}{M_{pl}}=(\frac{27}{8})^{\frac{1}{4}}m(\beta )^{-\frac{1}{4}%
}(1+2\beta m(\beta ))^{\frac{1}{2}}(\Delta _{\cal
R}^{2})^{\frac{1}{4}}. \label{ending3}
\end{equation}
Substituting the above equation into Eq. (\ref{epsilon2}), we have
the following expression for $\epsilon _{v}$:
\begin{equation}
\epsilon
_{v}=\frac{1}{48}\left(\frac{3}{2}\right)^{\frac{1}{2}}(\Delta
_{\cal R}^{2})^{\frac{1}{2}}m(\beta )^{-\frac{5}{2}}(1+2\beta
m(\beta ))^{3}. \label{epsilon3}
\end{equation}
The WMAP five-year data give the amplitude of the primordial scalar
power spectrum, $\Delta _{\cal R}^{2}\simeq 2.4\times 10^{-9}$ for
$N\sim 50$ \cite{Komastu08,Lambda}. Thus, so far, all of the slow
roll parameters can be expressed as the functions of $\beta$ (and
$N$). The spectral index and its running are
\begin{equation}
n_{s}=1-6\epsilon _{v}+2\eta _{v},~~~\alpha _{s}=-24\epsilon
_{v}^{2}+16\epsilon _{v}\eta _{v}-2\xi _{v},  \label{power2}
\end{equation}%
and the tensor-to-scalar ratio is
\begin{equation}
r=16\epsilon _{v}.  \label{tensor}
\end{equation}
\subsection{What can observables tell us?}
Inflation models can predict some observables, such as the spectral
index, its running, and the tensor-to-scalar ratio, etc., from which
one can link the observations to the theory. In particular, the
brane inflation models also predict the existence of cosmic
superstrings that open a significant window for testing the
superstring theory. In this subsection, we shall discuss how these
observables provide constraints on the parameter $\beta$ of the
KKLMMT model.
\subsubsection{Tensor-to-scalar ratio}
\begin{figure}[tbh]
\centerline{\includegraphics[bb=0 0 475 308,
width=3.5in]{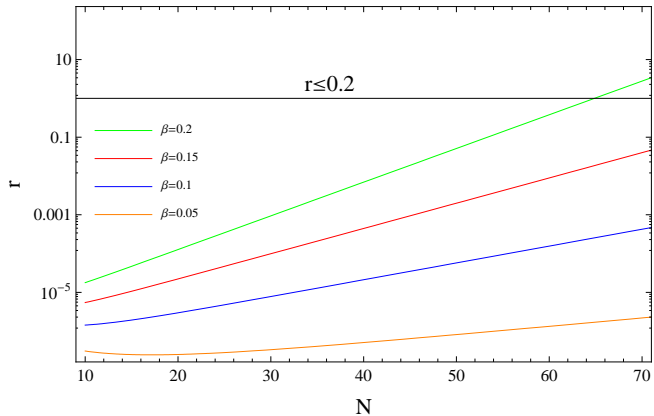}} \caption{The tensor-to-scalar ratio in
the KKLMMT model.} \label{tensor1}
\end{figure}

First, let's discuss the tensor-to-scalar ratio $r$. The
tensor-to-scalar ratio is an important observable that can
distinguish the inflationary cosmology from other scenarios. Also,
it can be used to distinguish different inflation models because
different models give distinctive predictions on this observable
quantity. Although the observations still have no ability to detect
the primordial gravitational waves in a convincing manner, an upper
limit could be given at least. Now we try to see how this observable
with such a loose upper-limit can set constraint on the KKLMMT
model. In Fig. \ref{tensor1}, we plot the tensor-to-scalar ratio $r$
vs. the $e$-folding number $N$ for the KKLMMT model, according to
Eq. (\ref{tensor}).

In Fig. \ref{tensor1}, we see that the larger value the parameter
$\beta$ takes, the greater value the tensor-to-scalar ratio will be.
This is due to the fact that the greater the value of $\beta$ is,
the faster the inflaton moves, so the larger tensor mode will be
produced in the inflationary era \cite{Baumann08}. Currently, the
tightest constraint on $r$ is $r<0.2$ (WMAP5+SNIa+BAO)
\cite{Komastu08,Lambda}. From Fig. \ref{tensor1} we see that the
green curve (corresponding to $\beta=0.2$) definitely violates this
bound, and we learn that the parameter $\beta$ cannot be greater
than $0.15$. Hence, we find that even such a loose bound on $r$
gives a so small value of $\beta$ about ${\cal O}(10^{-1})$, then
the true value of $\beta$ must be fine-tuned to some extent. We hope
the future CMB experiments could further constrain $r$ toward
$0.01$, providing true test on the KKLMMT model.
\begin{figure}[tbh]
\centerline{\includegraphics[bb=0 0 454
302,width=3.3in]{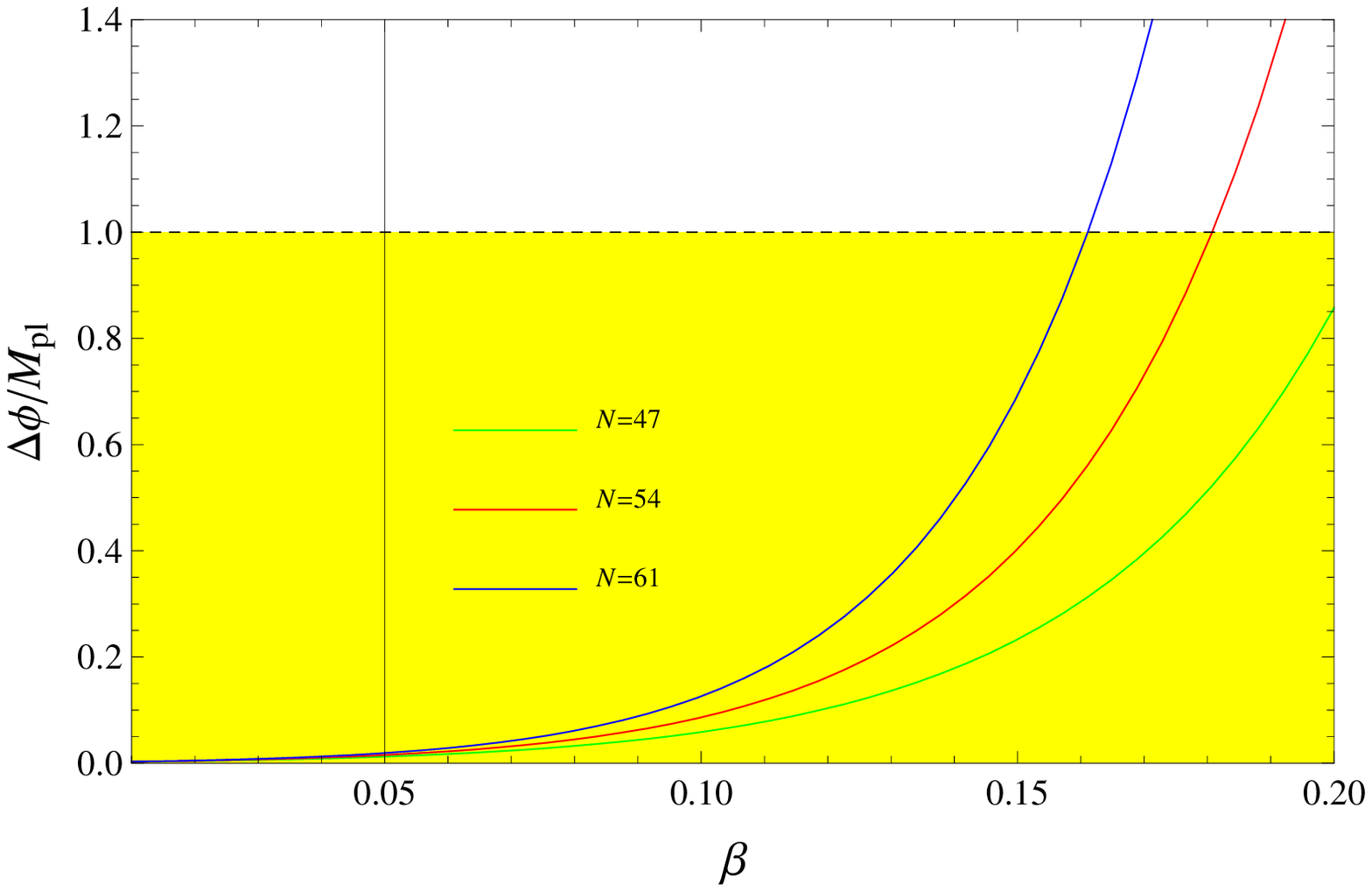}
\includegraphics[bb=0 0 348 272,
width=3.3in]{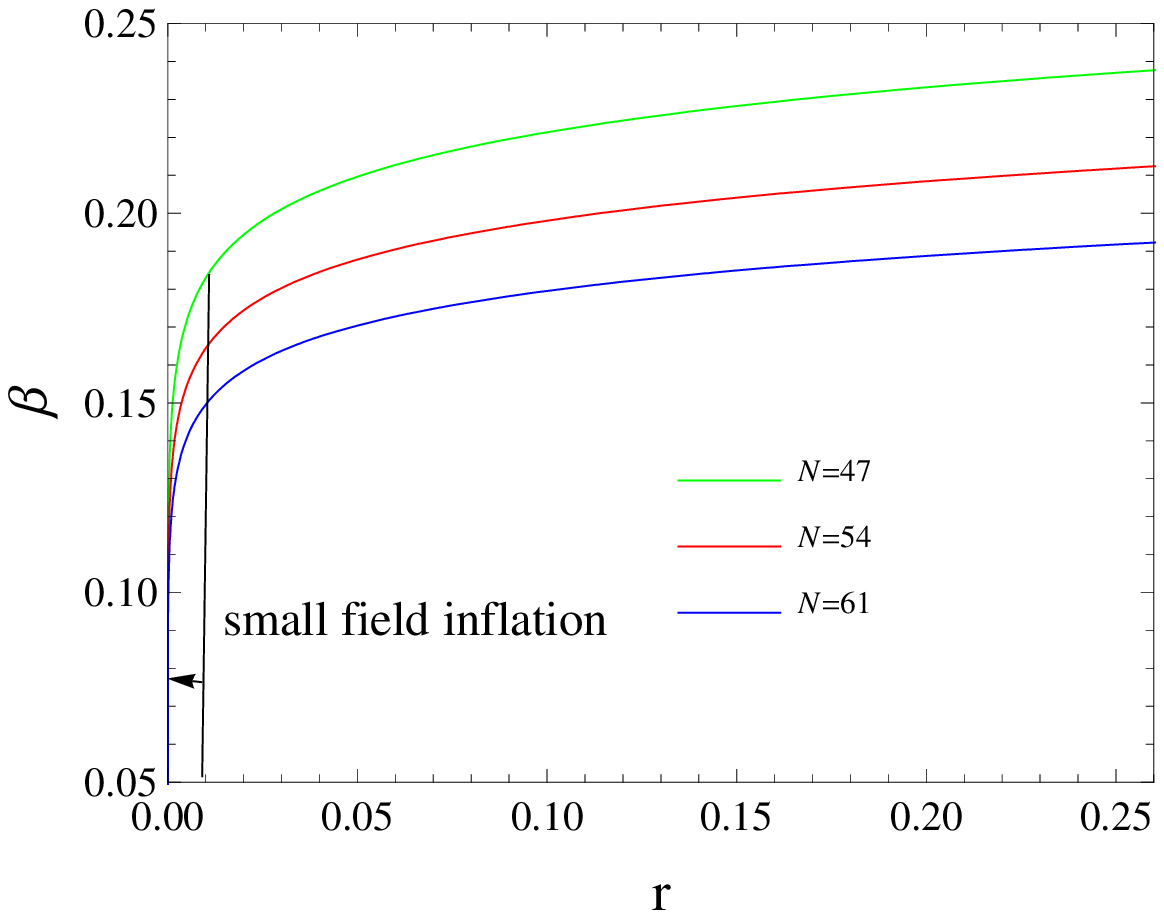}} \caption{\small Left: the change
magnitude of the inflaton field vs. the parameter $\beta$, for
different number of e-folds $N$; Right: the parameter $\beta$ needed
for the small $r$ value.} \label{field1}
\end{figure}

The above constraint on $\beta$ can also be examined from the field
value perspective. We can calculate the change magnitude of the
inflaton field from the moment when the fluctuations were generated
to the end of inflation,
\begin{eqnarray}
\frac{\Delta \phi }{M_{pl}} &=&\frac{\phi _{N}-\phi _{f}}{M_{pl}}  \notag \\
&=&\frac{3^{\frac{7}{12}}}{2^{\frac{5}{4}}}m(\beta )^{-\frac{5}{12}
}(1+2\beta m(\beta ))^{\frac{1}{2}}(\Delta _{\cal
R}^{2})^{\frac{1}{4}}\left[(24m(\beta
))^{{1/6}}-\left(\frac{20}{1+{\beta /3}}\right)^{{1/6}}\right].
\end{eqnarray}
According to \cite{Lyth96}, for a single field inflation model, the
change magnitude of the field has a relationship with the
tensor-to-scale ratio as
\begin{equation}
\frac{\Delta \phi }{M_{pl}}\geq
\left(\frac{r}{0.01}\right)^{\frac{1}{2}}. \label{field and tensor}
\end{equation}
We plot the change magnitude of the field as a function of $\beta $
for different e-folding number $N$ in the left panel of Fig.
\ref{field1}. It is very explicit that if $\beta \geq 0.2$, the
change magnitude of the field will be definitely greater than the
planck mass, which might not be consistent with the conditions of
effective field theory \cite{ArkaniHamed:2006dz,Huang:07}. The right
panel of Fig. \ref {field1} shows the allowed values for the
parameter $\beta$. According to Eq. (\ref{field and tensor}), if the
change of the field during the whole period of inflation is less
than the Planck mass, the tensor-to-scalar ratio will be surely less
than $0.01$. Thus, from right panel of Fig. \ref {field1}, it can be
inferred that $\beta$ should be less than $0.17$. Therefore, to
satisfy the current constraint on the tensor-to-scalar ratio and the
physical bound on the change magnitude of the field value, $\Delta
\phi \leq M_{pl}$, the parameter $\beta$ could not be greater than
$0.15$, which is both an observational and a physical bound on the
KKLMMT model.
\subsubsection{Cosmic string tension}
Cosmic strings and other topological defects have long been proposed
as one of the candidates for the origin of structure formation
\cite{Zeldovich80}, however, it has been shown that this scenario
leads to predictions incompatible with observations such as the
power spectrum of cosmic microwave background (CMB) temperature
anisotropy \cite{Pen97}. Nevertheless, the brane/anti-brane
inflation scenario inspired from string theory naturally indicates
that cosmic strings would have a small contribution to the CMB,
which is compatible with the observational limits
\cite{Jones02,Sarangi02,Zhang06}. This has led to a significant
revival of all aspects of cosmic string scenario, including new
theoretical motivations, phenomenological implications and direct
observational searches, see e.g.
\cite{Tye05,Copeland:2003bj,Jackson:2004zg,Shandera:2006ax,Wyman05}.
The possibility of detecting the signal of the cosmic strings
through astronomical observations opens a significant window to test
string theory. The current observational bound (the upper limit) for
cosmic string tension from the CMB temperature is roughly
\cite{Wyman05}
\begin{equation}
G\mu _{\mathrm{obs}}\lesssim 1.8\times 10^{-7}\text{ }(1\sigma
\text{ CL}), \label{tensor1sigma}
\end{equation}
and
\begin{equation}
G\mu _{\mathrm{obs}}\lesssim 2.7\times 10^{-7}\text{ (}2\sigma
\text{ CL)}. \label{tensor2sigma}
\end{equation}
Obviously, the KKLMMT scenario predicts that the tension of these
cosmic strings is very small and their contribution to CMB is well
below the current limits \cite{Tye05,Shandera:2006ax}. However, now,
we try to do this inversely, i.e., we intend to see how this
observational bound puts a constraint on the parameter $\beta$. This
may show what the loosest observational result can tell us.

The $D3$-brane collides with the $\bar{D}3$-brane at the end of
inflation at the bottom of the throat. The annihilation of the
D3-$\overline{\rm D}3$-branes initiates the hot big bang epoch,
meanwhile D1-branes (i.e. D-strings) and fundamental closed strings
(i.e. F-strings) are also produced. The quantities of interest are
the D-string tension $G\mu_D$ and the F-string tension $G\mu_F$,
where $\mu_D$ and $\mu_F$ are the effective tensions measured from
the viewpoint of the four dimensional effective action. Since in ten
dimensions, there is the relationship $T_3=T_F^2/2\pi
g_s=T_D^2g_s/2\pi$, where $T_3=1/(2\pi)^3g_s\alpha'^2$ is the
D3-brane tension, we have the tensions of the type IIB string in the
inflationary throat
\begin{equation}
G\mu_{F}=GT_{F}h^2=\sqrt{\frac{g_s}{32\pi}}\left({\frac{T_3h^4}{M_{\mathrm{pl}}^4
}}\right)^{1/2},  \label{Ftension}
\end{equation}
\begin{equation}
G\mu_{D}=GT_{D}h^2=\sqrt{\frac{1}{32\pi
g_s}}\left({\frac{T_3h^4}{M_{\mathrm{pl} }^4}}\right)^{1/2},
\label{Dtension}
\end{equation}
where $g_s$ is the string coupling. There are also bound states of
$p$ F-strings and $q$ D-strings with a cosmic string network
spectrum \cite{Copeland:2003bj}
\begin{equation}
\mu_{(p,q)}=\mu_F\sqrt{p^2+q^2/g_s^2}.
\end{equation}
In Eqs. (\ref{Ftension}) and (\ref{Dtension}) both F-strings and
D-strings are dependent on the string coupling $g_s$, but the
geometric mean $(\mu_F\mu_D)^{1/2}$ is independent of $g_s$. Thus,
we can define \cite{Zhang06}
\begin{equation}
G\mu_{s}=G(\mu_{F}\mu_{D})^{1/2}=\sqrt{\frac{1}{32\pi}}\left({\frac{T_3h^4}{M_{%
\mathrm{pl}}^4}}\right)^{1/2}. \label{stension1}
\end{equation}
Obviously, for F-string we have $\mu_F=\mu_{s}\sqrt{g_s}$, and for
D-string we have $\mu_D=\mu_{s}/\sqrt{g_s}$. The value of $g_s$ is
likely in the range 0.1 to 1 \cite{Copeland:2003bj}. Note that
$g_s>1$ can be converted to $g_s<1$ by $S$-duality.

Then, combining with COBE normalization (\ref{amplitude}), we obtain
the cosmic string tension in the KKLMMT inflation model
\begin{eqnarray}
G\mu _{s}=\frac{1}{18}(\frac{\pi }{3})^{\frac{1}{2}}(\frac{27}{8})^{\frac{3}{4}%
}m(\beta )^{-\frac{5}{4}}(1+2\beta m(\beta ))^{\frac{3}{2}}(\Delta
_{\cal R}^{2})^{\frac{3}{4}}. \label{stension2}
\end{eqnarray}

\begin{figure}[tbh]
\centerline{\includegraphics[bb=0 0 390
394,width=3.5in]{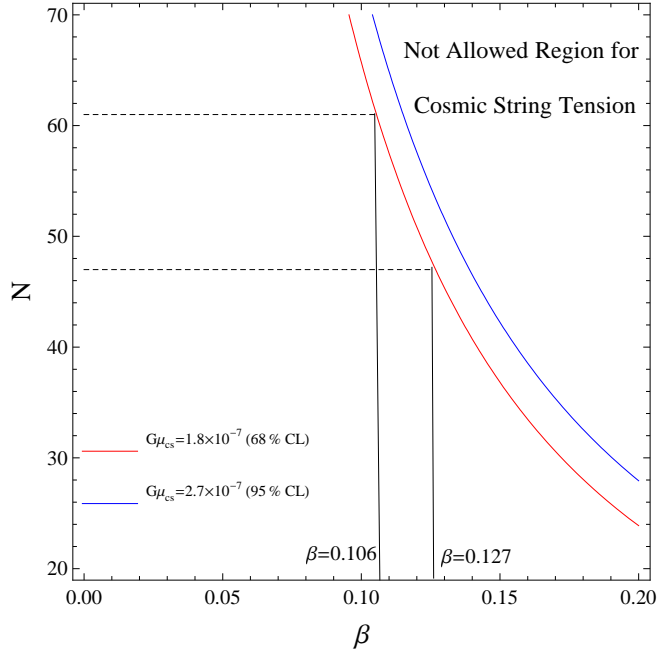}} \caption{Constraints from the
observational bound of the cosmic string tension.}
\label{figtension}
\end{figure}

Equation (\ref{stension2}) shows that the cosmic string tension
$G\mu_{s}$ depends both on the parameter $\beta$ and the number of
e-folds $N$. Thus, in order to see how the cosmic string tension
relies on the two parameters, we plot the correlation of $N$-$\beta$
according to the cosmic string tension $G\mu_{s}$ in Fig.
\ref{figtension}.

In Fig. \ref{figtension}, the upper-right corner is the not-allowed
parameter region, since it represents a too large tension and it is
excluded outside the $2\sigma$ CL. The smaller value the cosmic
string tension takes, the deeper the curve will move toward the
lower-left corner, which indicates a smaller parameter $\beta$.
According to the $1\sigma$ bound (the red curve in Fig.
\ref{figtension}), if one takes $N=61$, the observational $1\sigma$
curve will give the tightest constraint on $\beta$, namely
$\beta\lesssim 0.106$; if taking $N=47$ , the bound on the parameter
$\beta$ from cosmic string tension will be looser, $\beta \lesssim
0.127$. Thus, roughly speaking, the current observational bound on
the cosmic string tension tells us the following information on the
parameter $\beta$, namely $\beta \lesssim 0.12$.
\subsubsection{Spectral index}
The spectral index is certainly the most important observable that
could distinguish different inflation models. The WMAP five-year
data and the combination of various measurements favor a red power
spectrum, which could put tight constraints on the KKLMMT model.
\begin{figure}[tbh]
\centerline{\includegraphics[bb=0 0 461 297,
width=3.3in]{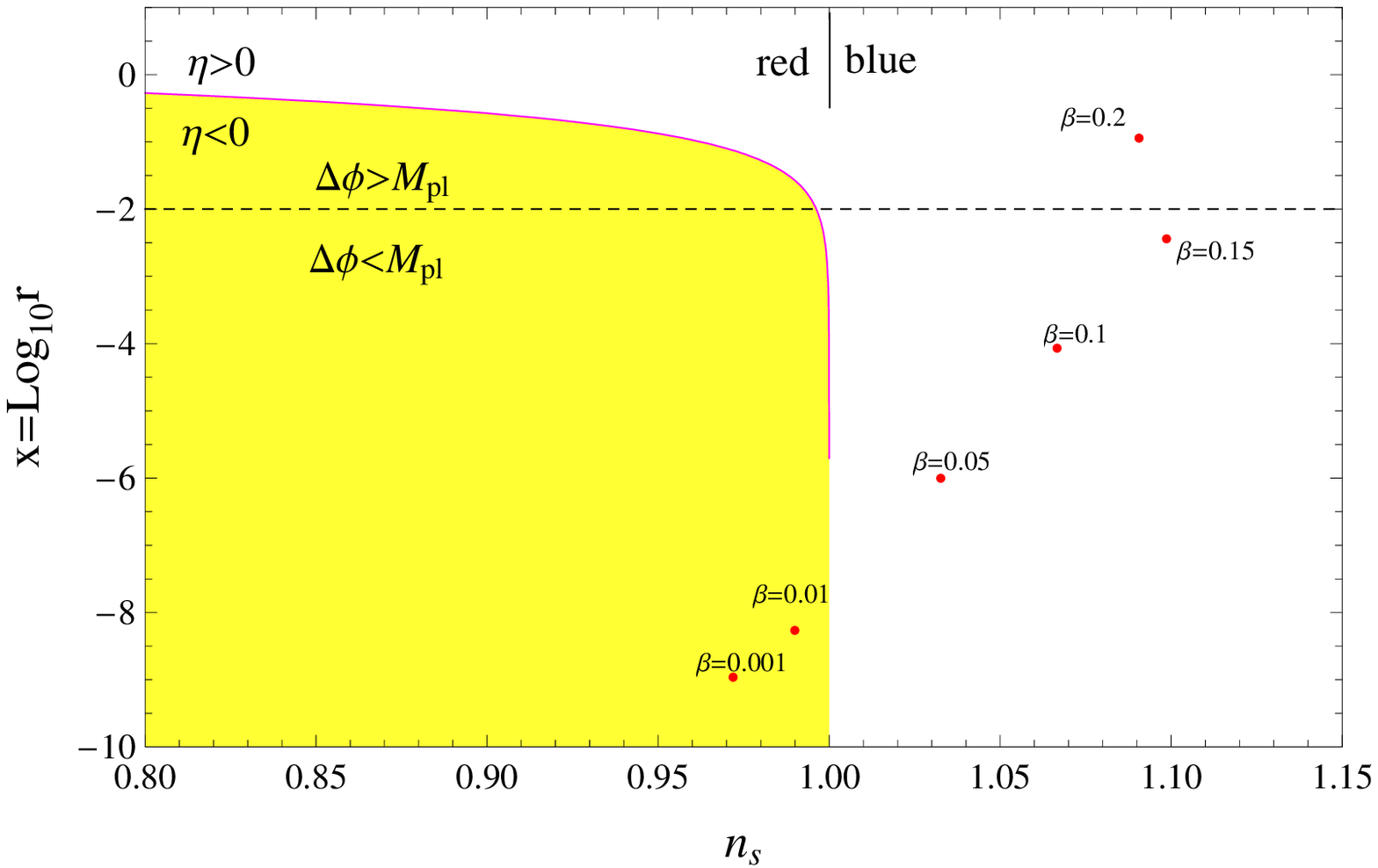}
\includegraphics[bb=0 0 381 233,width=3.3in]{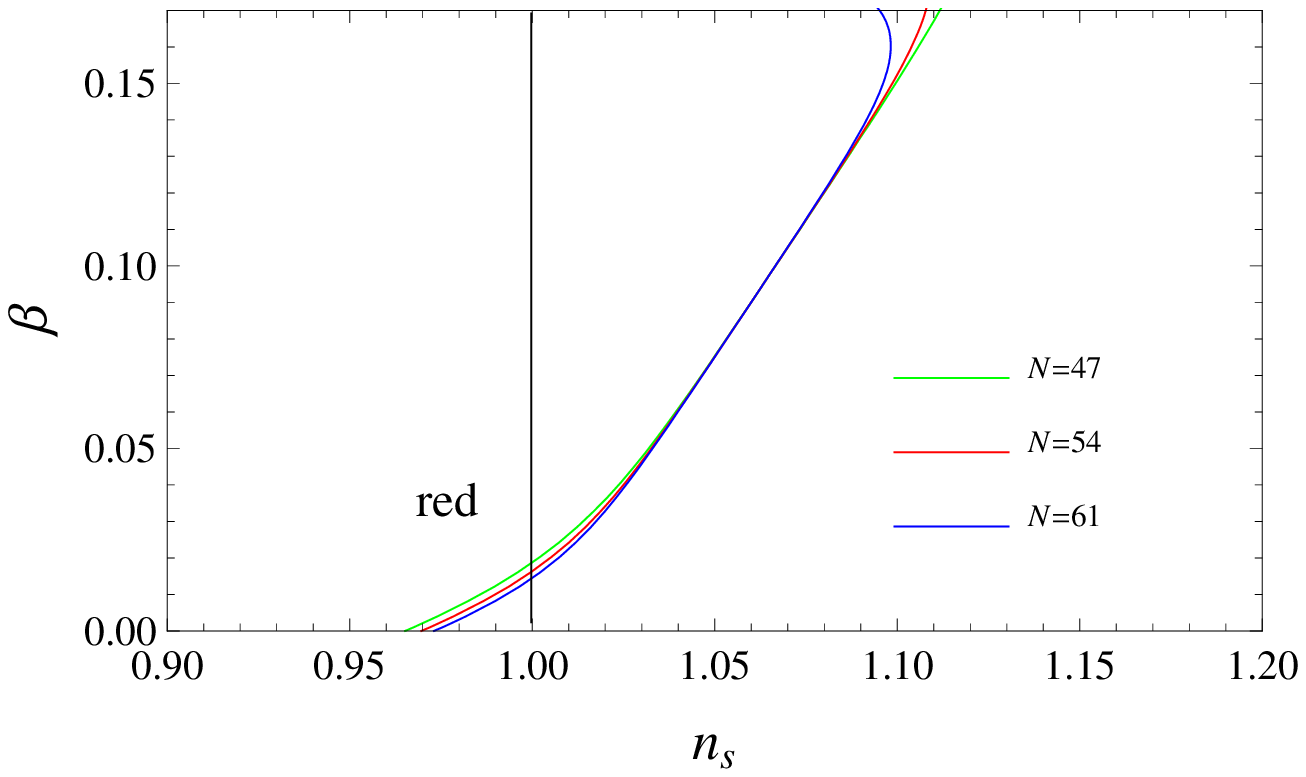}}
\caption{Left: Constraints on the KKLMMT model in the $r-n_s$ plane.
Right: $\beta$ vs. $n_s$ for different number of e-folds $N$.}
\label{model3}
\end{figure}

Now we plot the correlation of $r-n_{s}$ in the left panel of Fig.
\ref{model3} to see what is the trajectory of the model when varying
the parameter $\beta$. In this figure, by decreasing the parameter
$\beta$, we show that the trajectory of the model goes from the
upper-right corner to the lower-left corner in the $r-n_{s}$ plane.
So, only $\beta \lesssim 0.01$ could give both the red-tilted scalar
spectrum and the small field manner. This is due to the fact that a
larger parameter $\beta$ will make the field rolling too fast, which
might make the slow-roll parameter $\eta$ be positive. Therefore,
the red-tilted power spectrum requires $\beta \lesssim 0.01$. We
also plot $\beta$ vs. $n_{s}$ for different number of e-folds $N$ in
the right panel of Fig. \ref{model3}. In this figure, it is easy to
see that different number of $e$-folds $N$ can make little influence
on the curves, and the red-tilted spectrum roughly requires $\beta
\lesssim 0.01$.

Summarizing the above discussions, we see that the spectral index
can put the tightest constraints on the KKLMMT model. This is
obvious because the spectral index is the very quantity that has
been precisely measured while other quantities only get loose upper
limits from the observations. The aim of making these discussions is
to see from various angles of view what constraints these
observational quantities can bring to the model. From the above
analysis, we get to know that the current observational bound on the
parameter $\beta$ is roughly
\begin{eqnarray}
\beta \lesssim {\cal O}(10^{-2}) .\label{betabound2}
\end{eqnarray}
Further constraints from the WMAP five-year results will be
discussed in the next subsection.


We now pause to discuss the physical effect of the parameter
$\beta$. If $\beta$ goes larger, the potential $V(\phi)$ will become
steeper, which leads the inflaton to roll faster than the small
$\beta$ case. Thus, the change magnitude of the inflaton field
during inflation will be greater. At the same time, the $D3$ and
$\bar{D}3$-branes will have greater momentum to collide, producing
more $D$-strings and $F$-strings. Therefore, constraining the
parameter $\beta$ is actually constraining the shape of the
potential and the dynamics of inflation.
\subsection{Comparing to the WMAP five-year results}
The WMAP three-year data give a rather red spectrum, i.e. $n_s=
0.951^{+0.015}_{-0.019}$ (1$\sigma$ CL) for WMAP3 data only, and
$n_s=0.948^{+0.015}_{-0.018}$ (1$\sigma$ CL) for the combination of
WMAP3 and SDSS data. Therefore, such a red spectrum requires the
parameter $\beta$ in the KKLMMT to be fine-tuned. In \cite{Huang06},
it has been shown that, with the WMAP three-year data, the
constraints on the parameter $\beta$ are $\beta\leq 6\times 10^{-4}$
at the level of 1$\sigma$ and $\beta\leq 8\times 10^{-3}$ at the
level of 2$\sigma$ for WMAP3 only; $\beta\leq 6\times 10^{-3}$ at
the level of 2$\sigma$ for WMAP3 + SDSS. So, the fine-tuning for the
parameter $\beta$ is needed, according to the constraints from the
WMAP three-year results (see also \cite{Zhang06}).

However, the scalar spectral index derived from the WMAP five-year
data is relatively blue comparing to that of WMAP3, though it is
still red-tilted. See Eqs. (\ref{nsnumber1}) and (\ref{nsnumber2})
for the results of WMAP5. Therefore, we expect that these results
could relax the fine-tuning of $\beta$ to some extent.
\begin{figure}[tbh]
\centerline{\includegraphics[bb=0 0 387 259,
width=3.3in]{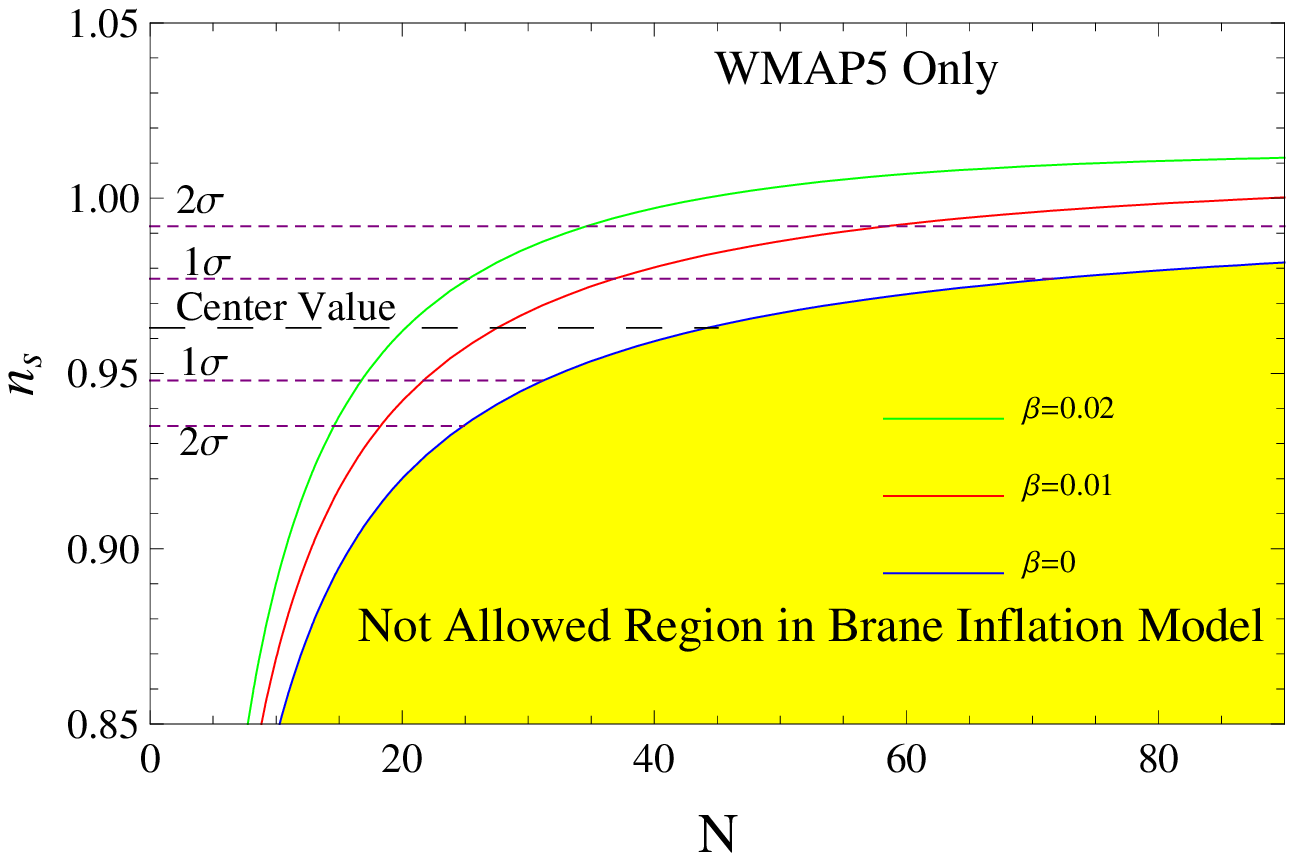}
\includegraphics[bb=0 0 391 258,width=3.3in]{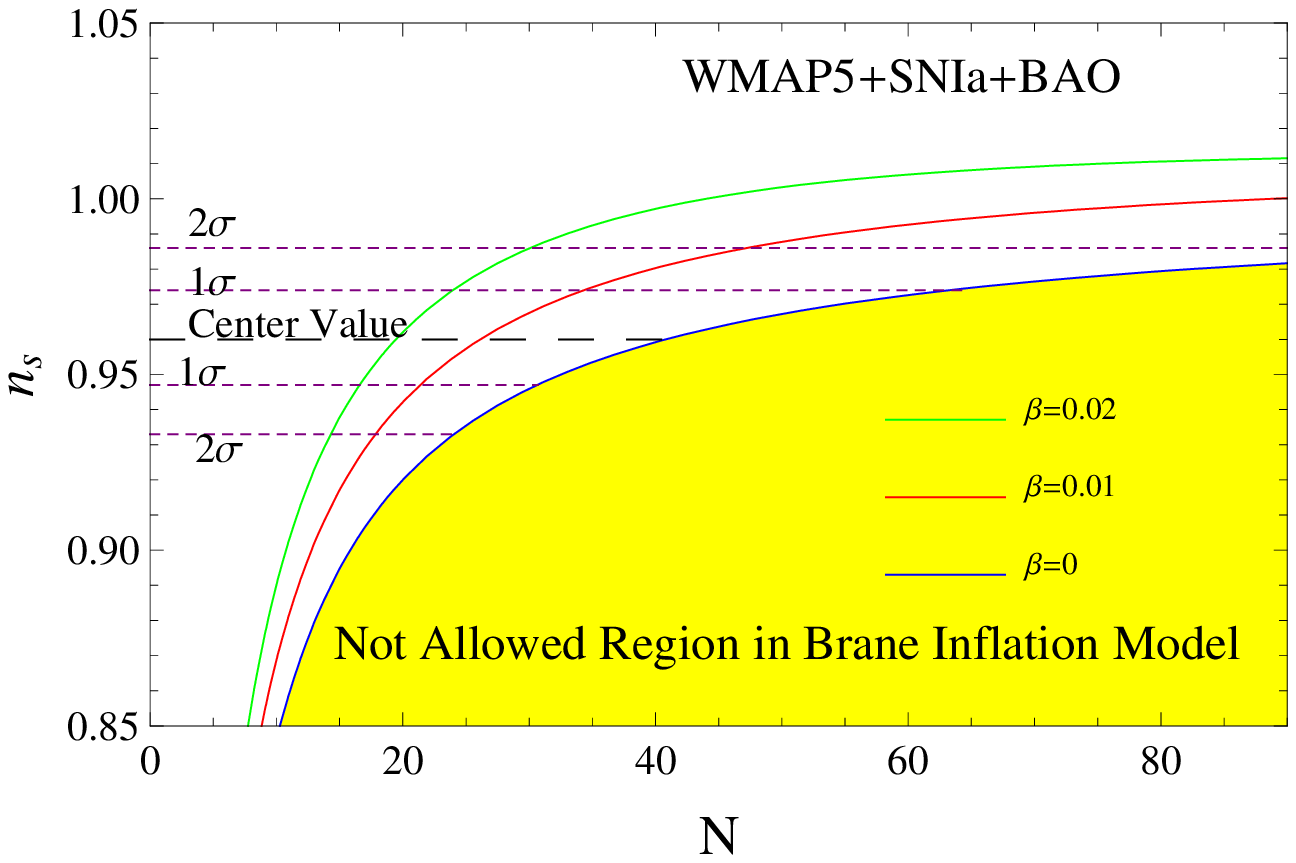}}
\caption{Comparison of the WMAP five-year results of the special
index with the KKLMMT model. Note that $\beta<0$ is forbidden since
in this case the $\beta$ term in the potential tends to push the
D3-brane out of the throat such that the inflation will not happen.
Left: WMAP five-year data only; right: WMAP five-year data combined
with SNIa and BAO data. } \label{ns2}
\end{figure}

We plot the spectral index $n_{s}$ as a function of the number of
$e$-folds $N$ in Fig. \ref{ns2}. For comparison, we also show the
observational results of the spectral index, from the WMAP five-year
data. In this figure, it is easy to see that the larger the
parameter $\beta$, the greater the spectral index. We also find that
the curves are very sensitive to the parameter $\beta$, i.e., the
curves of the spectral index will change significantly even though
the parameter $\beta$ only changes a little, e.g., from $0$ to
$0.01$. Note that $\beta<0$ is forbidden since in this case the
$\beta$ term in the potential tends to push the D3-brane out of the
throat such that the inflation will not happen. For the boundary
curve corresponding to $\beta=0$, the central value of $n_s$ gives a
reasonable value of the $e$-folding number $N$; however, when
changing $\beta$ to $0.01$, the model could not provide the central
value of $n_{s}$ within its reasonable range of the number of
$e$-folds, which implies that the parameter $\beta$ tends to be
fine-tuned confronting a red-tilted spectrum of the primordial
perturbation.
\begin{figure}[htbp]
\centering
\begin{center}
$\begin{array}{c@{\hspace{0.2in}}c} \multicolumn{1}{l}{\mbox{}} &
\multicolumn{1}{l}{\mbox{}} \\
\includegraphics[scale=0.85]{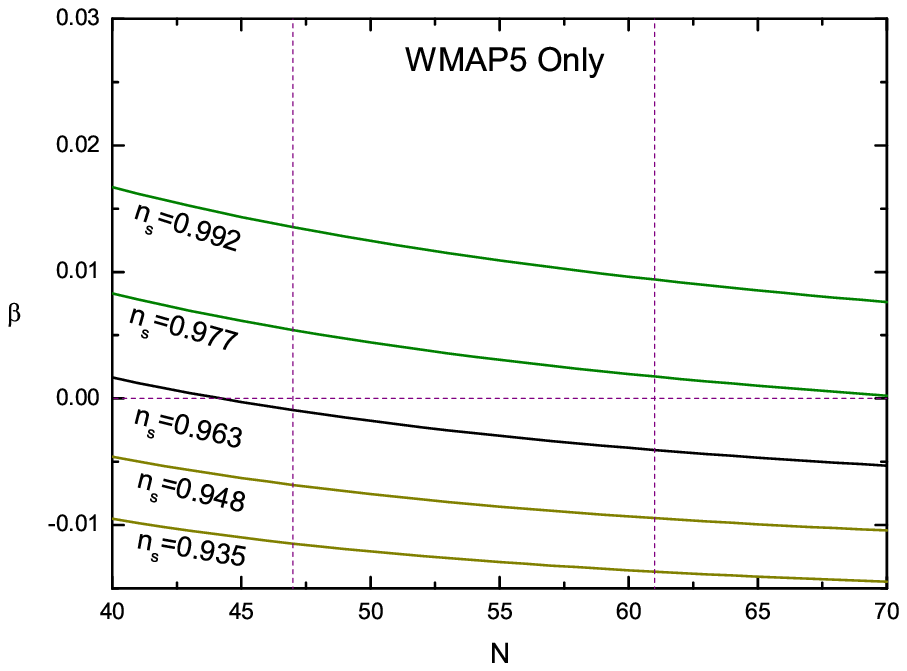} &\includegraphics[scale=0.85]{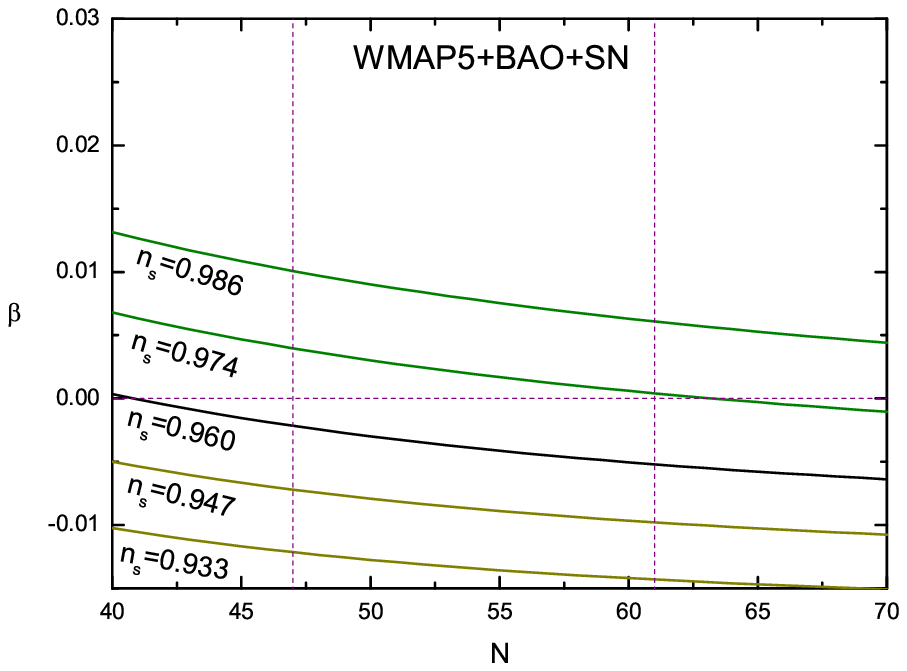} \\
\end{array}$
\end{center}
\caption[]{\small The constraint relationship between the parameter
$\beta$ and the $e$-folding number $N$ in the KKLMMT model,
according to the WMAP five-year results of the spectral index of the
primordial density perturbation (within 2 standard deviations). Note
that the area of $\beta>0$ and $47<N<61$ is a rational region for
the KKLMMT inflation.}\label{fig:nbeta}
\end{figure}

Confronting the WMAP five-year results of the spectral index, Eqs.
(\ref{nsnumber1}) and (\ref{nsnumber2}), the correlation between the
parameter $\beta$ and the $e$-folding number $N$ can also be
derived, see Fig. \ref{fig:nbeta}. From this figure, one can clearly
read off the allowed region in the parameter-space by the WMAP
five-year data. Note that the area of $\beta>0$ and $47<N<61$ is a
rational region for the KKLMMT model. Therefore, according to the
WMAP five-year results of the spectral index, we see that the KKLMMT
model can fit well with both the data of WMAP only and WMAP + BAO +
SN within 1$\sigma$ range. For WMAP only, we derive $\beta\leq
5.4\times 10^{-3}$ at the level of 1$\sigma$ and $\beta\leq
1.4\times 10^{-2}$ at the level of 2$\sigma$. For WMAP + BAO + SN,
we derive $\beta\leq 4.0\times 10^{-3}$ at the level of 1$\sigma$
and $\beta\leq 1.0\times 10^{-2}$ at the level of 2$\sigma$. So,
comparing to the WMAP five-year results, we find that the value of
the parameter $\beta$ is relaxed to ${\cal O}(10^{-2})$ at the level
of 2$\sigma$. The problem of fine-tuning of $\beta$ is alleviated to
a certain extent when confronting the WMAP five-year data. This is,
without doubt, a good news for the KKLMMT model.
\begin{figure}[htbp]
\begin{center}
\includegraphics[scale=0.6]{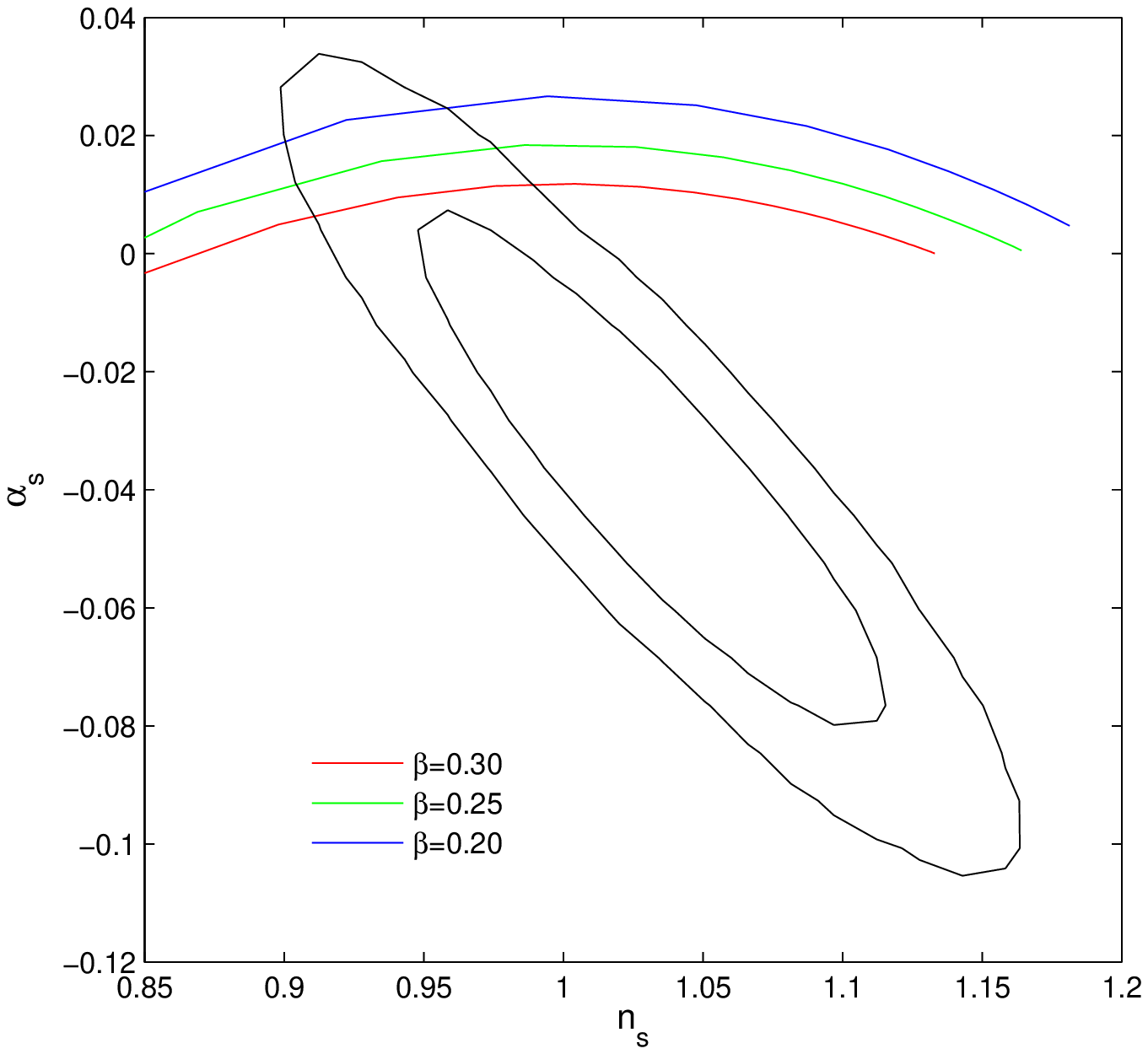}
\caption[]{\small Comparing the KKLMMT model to the WMAP five-year
data (WMAP5 only): the running of the spectral
index.}\label{fig:running}
\end{center}
\end{figure}

For the running of the spectral index, according to the WMAP
five-year data, one finds an upward shift from the three-year
result, $\alpha_s=-0.032^{+0.021}_{-0.020}$ (WMAP5 + BAO +
SN).\footnote{Such a small value, $\alpha_s\sim -{\cal O}(10^{-2})$,
may indicate that there is no evidence for the running of the
spectral index, as pointed out in \cite{Komastu08}.} The KKLMMT
model, generally, cannot yield a seeable running of the spectral
index. However, for $\beta>0.1$, a large running is also possible in
the KKLMMT model. Unfortunately, even though taking a big $\beta$,
the KKLMMT model cannot provide a running compatible with the
current WMAP result for the running, see Fig. \ref{fig:running}. Of
course, the future high precision observations might make a more
precise measure for the running of the spectral index, and might
demonstrate explicitly that there is no running of the spectral
index.

Furthermore, we also compare the angular power spectrum of the
KKLMMT model with the WMAP five-year data, by employing the CAMB
code \cite{CAMB}, see Fig. \ref{angular}. In the right panel of Fig.
\ref{angular}, we see that the smaller the value of $\beta$ is, the
larger the first peak amplitude will be, and, obviously, $\beta=0.1$
is not able to fit the power spectrum at all. Thus, the current data
of the CMB angular power spectrum also indicate that the parameter
$\beta$ should be taken as $\lesssim 0.01$, which is consistent with
the constraint from the spectral index.

Finally, let's discuss the prospects for the future measurements of
the tensor-to-scalar ratio, $r$, in light of the existing constraint
on $n_s$. From the above analysis, we know that the current
constraints on $n_s$ put a bound on the KKLMMT model of order
$\beta<0.01$ or so. Such a small value of $\beta$ leads to a
negligible $r$, less than ${\cal O}(10^{-5})$, see Fig.
\ref{tensor1}. This means that if current bounds on $n_s$ do not
shift appreciably, any observation of primordial tensor modes will
rule out the KKLMMT model entirely. Therefore, the future
measurements of $r$ might provide a disproof for the KKLMMT model.

\section{Conclusion}
In this paper, we have studied brane inflation with the WMAP
five-year data. We first considered a general brane inflation model
in which the problem of dynamic stabilization is neglected.
Furthermore, we examined the KKLMMT model by using the WMAP
five-year results.

For the general inflation model, we show that, according to the WMAP
five-year data, the model survives at the level of 1$\sigma$. This
conclusion is different from that of a previous work \cite{Huang06}
which is based on the analysis of the WMAP three-year data. In
\cite{Huang06}, with the WMAP three-year data, it was indicated that
the brane inflation model for $n=4$ (D3-brane case) is near the
boundary of the WMAP3 only and cannot fit the WMAP3+SDSS data at the
level of 1$\sigma$; the model with $n=2$ (D5-brane case) is outside
the range allowed by WMAP3 only or WMAP3+SDSS at the level of
1$\sigma$. Therefore, we find that the WMAP five-year data save the
brane inflation model to the level of $1\sigma$.

For the KKLMMT model, we first discuss how the various observables
can bring constraints on the parameter $\beta$ of the model. Then,
we compare the model to the WMAP five-year results. In
\cite{Huang06}, by using the WMAP three-year data, the authors find
that the KKLMMT model cannot fit WMAP3+SDSS data at the level of
1$\sigma$ and a fine-tuning, at least eight parts in a thousand, is
needed at the level of 2$\sigma$. However, in this work, we show
that the KKLMMT model is consistent with both the data of WMAP5 only
and of WMAP5+BAO+SN at the level of $1\sigma$. Moreover, comparing
to the WMAP five-year results, we find that the value of the
parameter $\beta$ is relaxed to ${\cal O}(10^{-2})$ at the level of
2$\sigma$. Therefore, the problem of fine-tuning of $\beta$ is
alleviated to a certain extent when confronting the WMAP five-year
data. This is definitely a good news for the KKLMMT model.

This paper only discusses the the simplest realistic brane
inflationary model, namely the KKLMMT model, in which the usual
slow-roll scenario is employed, but does not involve the
Dirac-Born-Infeld (DBI) scenario, in which the rolling of the
inflaton is albeit slow but relativistic. In the DBI model, large
tensor mode and/or non-Gaussianity may emerge, providing a possible
stringy signature. The DBI model has been investigated in detail in
\cite{Alabidi:2008ej} by using the WMAP five-year data.

\section*{Acknowledgments}
One of us (YZM) would like to thank  Anne-Christine Davis, George
Efstathiou, Hiranya Peiris, Qing-Guo Huang, Simeon Bird and Yi Wang
for helpful discussions, also thank Duncan Hanson and Damien Quinn
for computer help. This work was supported by grants from the
Cambridge Overseas Trust and the studentship from Trinity College,
Cambridge, and the Natural Science Foundation of China.

\end{document}